\documentclass{elsart}

\usepackage{graphicx}
\usepackage{amssymb}
\usepackage{harvard}

\begin{document}

\begin{frontmatter}



\newcommand{\3}{\ss}
\newcommand{\n}{\noindent}
\newcommand{\eps}{\varepsilon}
\newcommand{\be}{\begin{equation}}
\newcommand{\ee}{\end{equation}}
\def\ba{\begin{eqnarray}}
\def\ea{\end{eqnarray}}
\def\de{\partial}
\def\msun{M_\odot}
\def\div{\nabla\cdot}
\def\grad{\nabla}
\def\rot{\nabla\times}
\def\ltsima{$\; \buildrel < \over \sim \;$}
\def\simlt{\lower.5ex\hbox{\ltsima}}
\def\gtsima{$\; \buildrel > \over \sim \;$}
\def\simgt{\lower.5ex\hbox{\gtsima}}

\title{Primordial gas cooling behind shock waves in merging halos}

\author[1,2]{E.~O.~Vasiliev}\ead{eugstar@mail.ru},
\author[3]{Yu.~A.~Shchekinov}\ead{yus@phys.rsu.ru}

\address[1]{Tartu Observatory, 61602 T\~oravere, Estonia}
\address[2]{Institute of Physics, University of Rostov,
Stachki St. 194, Rostov-on-Don, 344090 Russia}
\address[3]{Department of Physics, University of Rostov,
Sorge St. 5, Rostov-on-Don, 344090 Russia}

\begin{abstract}
We investigate thermal regime of the baryons behind shock waves arising
in the process of virialization of dark matter halos. We find a fraction of
the shocked gas cooled by radiation of HD molecules down to the temperature
of the cosmic microwave background (CMB): this fraction increases sharply
from about $f_{\rm c}\sim 10^{-3}$ for dark halos of $M=5\times 10^7\msun$ to
$\sim 0.1$ for halos with $M=10^8\msun$ at $z=10$.
We show, however, that further
increase of the mass does not lead to a significant growth of $f_{\rm c}$ --
the assymptotic value for $M\gg 10^8\msun$ is of 0.2. We estimate star formation 
rate associated with such shock waves, and show that it can be a small but 
not negligible fraction of the star formation connected with cooling by 
HI and H$_2$. We argue that extremely metal-poor low-mass stars in the 
Milky Way may have been formed from primordial gas behind such shocks. 

\end{abstract}


\end{frontmatter}

\section{Introduction}

It is widely believed that molecular hydrogen H$_2$ and its deuterated analogue
HD determine thermodynamics of primordial gas and characteristics
of the first stars \cite{lepp83,sh86,puy93,galli95,galli98,galli02,stancil98,tegmark97,galli97,puy97,puy98,uehara00,flower02,nakamura02,flower02,machida05,omukaiHD}.
In turn, the amount of H$_2$ and HD and their cooling efficacy greately depend on
dynamical and thermal regime of the gas. In particular, shock waves are
argued to strongly enhance the rate of conversion of atomic
hydrogen to its molecular form  
\cite{shent,sse83,shap87,shap92,sh91,ferrara,yamada,uehara00,cen,machida05,sv05,bromm05}.
On the other hand, first stars have formed from the gas which most likely was processed
through shock waves inevitably emerged during virialization of dark matter halos
\cite{shap93,haiman96,tegmark97,abel00,abel02}, and 
therefore possible enhancement of H$_2$ and
HD formation in these conditions can have important consequences for characteristics
of the first stars \cite{svaat,ohhaiman,sv03,sv05,bromm05}.

When dark matter halos form in the hierarchy of mergings of small mass
minihalos (see review in \cite{barkana01,ciardi04}),
shocks form and compress
the baryons. At sufficiently large velocities of colliding flows
($v>8$ km s$^{-1}$) fractional ionization in shocked gas increases above
the frozen cosmological value, and results in acceleration of chemical
kinetics of H$_2$ catalized by electrons. Moreover, collisions with velocities
above the critical value $v>8.6[(1+z)/20]^{-1/6}$ km s$^{-1}$ lead to a
rapid formation of HD and an efficient cooling down to the minimun
temperature $T=T_{\rm CMB}=2.7(1+z)$ \cite{sv05,bromm05,vs05,vs06}. The
latter result is inferred from simplified calculations of a Lagrangian
fluid element behind the shock, and can only show principal possibility
of the shocked gas to undergo an extreme cooling. 
Within a 1D code Ripamonti (2006) studied contribution of HD cooling 
in a contracting halo starting from an already virialized state 
free of molecular content, and found that a restricted region 
with baryon mass $\sim 200M_\odot$, 
where HD cooling is efficient can form even in low-mass halos, 
$M_h \sim 10^5M_\odot$ at $z = 20$, $v \simeq 3$~km~s$^{-1}$.

Explicit answer of how
big is a fraction of cooled gas can be found only in hydrodynamic
simulations. Full 3D simulations are expensive and time consuming.
For these reasons they always are made within a specified realization of a
random hydrodynamic field, and therefore represent only very restricted 
regions in the whole
space of possible random hydrodynamic fields of a given spectrum.
This therefore restricts the final thermodynamic
state of baryons within a range corresponding to chosen interelations
between the amplitudes of different wave modes. As a result, estimates of the
fraction of cold baryons able to form stars are biased by such limitations.
From this point of view
1D hydrodynamical simulations of the chemistry and thermodynamics of a
shocked primordial gas are important for a qualitative picture of what we can
expect in principle in the conditions preceeding formation of the first stars.
In this paper we show 1D computations of chemical and thermal regime of the
gas behind shock waves after a head-on collision of two clouds of equal sizes.
As was pointed by \cite{gilden}, in supresonic cloud collisions the
rarefaction time transverse to the symmetry axis is longer than the
crossing time $t_c=3R/2v_c$ -- the time for the shock to pass through the
entire cloud.
In calculations we associate the collisional velocities of the clouds with
the mass of the halo $M$ formed in this collision
\begin{equation}
\label{vel}
v_c=(3^4\pi^3\Omega_m\rho_c)^{1/6}G^{1/2}M^{1/3}(1+z)^{1/2},
\end{equation}
$\Omega_m$ is the matter closure parameter, $\rho_c=3H_0^2/8\pi G$ is the
critical density. We assume therefore that the minihalos move with the relative
velocity equal to the virial velocity of the larger halo.

In Section 2 we describe dominant thermo-chemical processes of a shocked gas;
Section 3 contains description of the postshock thermodynamics as 1D
calculations predict; in Section 4 we discuss consequences for formation
of the first stars; the discussion and summary of the results
are given in Section 5.

Throughout the paper we assume a $\Lambda$CDM cosmology with the parameters
$(\Omega_0, \Omega_{\Lambda}, \Omega_m, \Omega_b, h ) =
(1.0,\ 0.7,\ 0.3,\ 0.045,\ 0.7 )$ 
and deuterium abundance $2.6\times 10^{-5}$,
consistent with the most recent measurements \cite{sperg06}.

\section{Chemistry and thermal regime behind the shock}

In the center of mass of the colliding baryon components of two merging
minihalos a discontinuity forms at the symmetry plane, and two shock waves
begin to move outward. We assume that collisionless dark matter components
occupy considerably bigger volume and neglect gravitational
forces on baryons. Therefore we describe propagation of the shock
by single-fluid hydrodynamic equations with radiative energy losses
appropriate for primordial plasma: Compton cooling, recombination and
bremsstrahlung radiation, collisional excitation of HI \cite{cen92},
H$_2$ \cite{mckee,pineau} and HD \cite{flower00,lipovka05}. Chemical and ionization
composition include a standard set of species: H, H$^+$, H$^-$, He,
He$^+$, He$^{++}$, H$_2$, H$_2^+$, D, D$^+$, D$^-$, HD, HD$^+$, $e$. The
corresponding rates are taken from \cite{galli98,stancil98}; 
the shock wave was computed in one crossing time $t_c$.
We assumed a ``top-hat'' initial baryonic distribution 
in colliding haloes with density equal to the virialized value
$18\pi^2 \Omega_b \rho_0 (1+z)^3$, while temperature is taken to be close 
to the cosmic microwave background (CMB) 
temperature $T_b=1.1 T_{\rm CMB}$. This corresponds to a
simplified picture when merging halos are already compressed to their
virial radii, but not virialized yet. On the other hand, such assumptions
about the initial thermal state of the colliding halos allow to better 
understand the role which shock compression of baryons plays in their ionization
and chemical state in the process of virialization.
The fractional ionization $x$, and the abundances of H$_2$ and HD
molecules before the shock are taken equal to their background values 
$x=10^{-4}$, $f({\rm H}_2)=10^{-5}$ and $f({\rm HD})=10^{-9}$, respectively. 
Note, that adjustment of baryons to gravitational potential of dark 
matter and the corresponding steepening of density profiles in dark matter 
haloes occurs on timescales shorter than the time between mergers. 
In these conditions merging haloes may have already strongly stratified 
temperature distribution under efficient cooling in H$_2$ and HD lines 
\cite{ripa}. Thermal structure of baryons behind the shocks in merging 
nonuniform haloes will be described in a separate paper. 

Fig. 1 shows typical distributions of temperature, fractional concentrations
of H$_2$ and HD and their contribution to the total cooling behind the
shock front at $t=(0.2,~0.6,~1)t_c$,
for the halos merged at $z=20$ with $v_c=22$ km s$^{-1}$ corresponding to
the total mass $M=1.9\times 10^7M_\odot$. Three qualitatively different cooling
regimes can be distinguished in the temperature profile: in the high temperature
range ($1500<T<7000$ K) excitation of ro-vibrational levels of H$_2$ dominates,
while in the intermediate range ($200<T<1500$ K) only rotational lines contribute
to the cooling, and in the lowest range ($T<200$ K) rotational cooling
from H$_2$ molecules exhausts and only HD rotations support cooling -- it is
seen in the lower panel from comparison of the relative
contributions of H$_2$ and HD cooling.

In this particular case a small fraction ($q\simeq 0.1$ by mass) of
the shocked baryons close to the symmetry plane
has cooled to the minimum possible value $T\simeq T_{\rm CMB}=2.7(1+z)$ due to
cooling in rotational lines of HD. In general, the fraction of compressed baryons
cooled down to a certain level depends on the relative velocities of the colliding
clouds $v_c$: the larger the collisional velocity $v_c$,
the stronger the gas compression after
cooling, and the higher the contribution from HD cooling \cite{sv05}.
Fig. 2 shows the fraction of baryons $q_T(M)$ contained in several temperature ranges
versus the halo mass:
$T<200$ K, $T<150$ K, $T<100$ K, and at $T\simeq T_{\rm CMB}$, at one crossing
time $t=t_c$.
It is seen that in the temperature range $T<200$ K cooled
by H$_2$ radiation, $q_T(M)$ is equal to 0.06
for halo mass $10^7 M_\odot$ and asymptotically
(at $M\gg 2\times 10^7 M_\odot$) approaches 0.5.
One should stress that compressed baryons can have temperature below
150 K only due to a dominant contribution
from HD cooling. It is readily seen that $q_T(M)$ in the lower temperature range
($T<100$ K, and at $T=T_{\bf cmb}$) is a very sharp function of
the halo mass: for instance, at redshift $z=20$ (Fig. 2a)
a two-fold increase of the mass from $10^7
M_\odot$ to $2\times 10^7M_\odot$ results in a two-order of magnitude
increase of $q(T_{\rm CMB})$ from $10^{-3}$ to 0.1. At higher masses the
dependence flattens and asymptotically in the limit $M\gg 2\times 10^7M_\odot$
approaches $0.2$. At lower redshifts gas density decreases and halo radius
increases, as a result the collision time $t_c$ becomes longer and
$q_T(M)$ shifts towards bigger halo
masses, approximately by factor of 5 as seen from Fig. 2b for a collision occured
at $z=10$.

\section{Star formation}

Baryons cooled below temperature $T<200$ K are normally thought to be able to
fragment and initiate star formation. In gas layers compressed by shock waves
gravitational instability of the cooled gas occurs naturally when the thickness
of the layer equals the Jeans length. In order to understand the range of
masses expected to form through the instability we applied
Gilden (1984) criterion for a shock-compressed gas which imply that:
{\it i)} the characteristic growth time is shorter
than the collision time, and {\it ii)} the critical wavelength is shorter
than the initial size of the clouds. The corresponding critical mass
$M_{\rm cr}$ depends on the average temperature and density in the layer
under consideration. Therefore, when halos merge with small relative velocities
(corresponding to lower halo masses),
smaller fraction of the compressed baryon mass cools down to sufficiently
low temperatures to form an unstable layer, while mergings with higher
relative velocities increase the fraction of gravitationally unstable baryons.
Fig. 3 shows dependence of the halo masses  with a given fraction of the
compressed baryons unstable in Gilden sense vs redshift $z$.
Each line is marked with symbols corresponding to a fraction of baryon mass
$f_c(M,z)$ unstable againts fragmentation: for instance, when halos with masses
corresponding to the upper line $2.6\times 10^7[(1+z)/20]^{-2.4} M_\odot$ merge, half of
their mass becomes compressed in a layer of a cold
gas with temperature $T<100$ K unstable in Gilden sense. The fraction of
mass unstable against fragmentation increases with the mass of merging halos,
approximately as $f_c(M,z)\simeq \exp[-(2.24/m_0)^{2.5}]$, where the halo mass is
taken in the form $M_{\rm h}=m_0[(1+z)/20]^{-2.4} 10^7M_\odot$.  At the latest
stages the unstable layers of $f_c(M,z)$ are dominated by HD cooling, so that
fragments being formed in these conditions can reach the minimum possible
temperature $\simeq T_{\rm CMB}$. The corresponding Jeans mass in the unstable
layer is $M_J \leq 2.3\times 10^3 M_\odot v_{10}^{-1} [(1+z)/20]^{1/2}$,
which is considerably smaller than the baryonic mass of this layer;
here $v_{10} = v_c/10$~km~s$^{-1}$. With accounting (\ref{vel})
the Jeans mass in the unstable layer is $M_J\leq
1.3\times 10^4M^{-1/3}M_\odot$. This means that when halos
with masses $M>10^4M_\odot$ merge more than one cold and dense
clouds in the unstable layer can form and give rise to formation
of stars.

Whether such fragments are the protostellar condensations evolving further
in a single massive star, or they are the nestles, where in the process of
sequential (hierarchical, \cite{hoyle}) or of a one-step \cite{nakamura99} 
fragmentation a cluster of less massive
stars is formed, depends sensitively on details of gravitational contraction
and thermal evolution of the fragments 
(see discussion in \cite{coppi,glover}). However, both
the mass of a central protostellar core in the former case, and the minimum
mass of the protostellar condensation in the latter case are determined by
the opaqueness of the contracting gas.
\cite{vs05} estimated this mass as
$M_J\sim 10^{-3}(1+z)^{3/2}M_\odot$, what gives for
$z=10-20$ a relatively low mass limit: $M_{\rm J}\sim (0.03-0.1)M_\odot$.
Therefore stars formed in such cold layers with a predominance of HD
cooling on later stages, are anticipated in general 
to be less massive than those expected
when thermodynamics of primordial gas is determined by 
H$_2$ cooling \cite{sv05}.

The fraction of baryons in the universe able to cool below $T=100$ K and form
presumably low-mass stars behind shock waves in mergings can be estimated as

\begin{equation}
\label{massint}
f_c=\left.\int\limits_{M_{\rm c}}^{M_{4\sigma}}MF(M)\,dM\right/
\int\limits_{M_{\rm min}}^{M_{4\sigma}}MF(M)\,dM,
\end{equation}
where $M_{\rm c}$ is the halo mass at a given redshift, where the fraction of
cold ($T\leq 100$ K) baryons is equal to 1\%, $F(M)=dN/dM$ is the Press-Schechter
mass function, $M_{\rm min}$, the minimum halo mass at a given redshift; Fig. 4
depicts $f_c$ versus redshift. For $M_{\rm min}$ we have taken $10^3M_\odot$,
$10^4M_\odot$ and $10^5_\odot$, however, only the last value seems meaningful,
because for lower halo masses their baryonic content is too small and
likely can be easily removed in tidal interactions. Independend on
$M_{\rm min}$ the total fraction of baryons in the universe
able to cool below $T=100$ K and form stars approaches $f_c\sim 0.1$ at $z=10$. 
For comparison we show in Fig. 4 
fraction of baryons in the universe cooled by 
H$_2$ molecules and hydrogen atoms with $M_{\rm c}$ corresponds 
to $T_{vir} = 400,~10^4$~K, for both cases 
we assume that in virialized halos 8\% of baryons can 
have temperature 200 K as calculated by \cite{abel98}, and integrate over 
the halo mass spectrum from $M_{\rm min}=10^4~M_\odot$.

High fraction of baryons cooled down $\leq 150$~K in collisions of 
large halos gives high luminosity in HD lines. Earlier studies 
\cite{sh86,sh91,kamaya,omukai05} mainly pay attention to the emission
of clusters of dense clouds formed in very large proto-galaxies, 
$M\geq 10^8M_\odot$. As it was shown in the previous section 
during head-on collisions of smaller halos, $M\sim 10^{7}M_\odot$, 
a big fraction of baryonic mass can cool down to the conditions, 
where the emission in HD lines is the main process of energy losses. 
The luminosity in HD lines is estimated as 
$L_{\rm HD}^{tot} \sim \Lambda_{\rm HD} M_{\rm HD} x_{\rm HD}/(\mu m_p)$, 
where $\Lambda_{\rm HD}$ is the cooling rate per molecule, 
$M_{\rm HD} = f_c (\Omega_b/\Omega_m) M_h$, $x_{\rm HD}$ is the abundance 
of HD. For the final phases ($t = t_c$) of the 
collision of halos with the total mass 
$1.9\times 10^7M_\odot$, when the density in a compressed layer is 
$n \sim 400$~cm$^{-3}$ (see Fig.~1), the expected
liminosity in HD lines can reach $\sim 5\times 10^{35}$~erg~s$^{-1}$. 
This is an upper limit for the luminosity from protogalaxies at 
the stages before their fragmentation. Further collapse and fragmentation 
increase the luminosity. This makes plausible detection
of radiation in HD rotational lines from forming 
protogalaxies \cite{kamaya}. 

Characteristic star formation time inside the formed
fragments is determined by the baryon density in the fragments
$t_{\rm sf}=t_J/\varepsilon$, where $t_J$ is the Jeans time corresponding to the
baryon density in cold layers $\rho_f$, $\varepsilon\ll 1$
is the star formation efficiency. Gas densities in cold layers when they fragment
can be found from the condition $\rho_iv_c^2=k\rho_fT_{\rm CMB}/m_H$, which gives
\begin{equation}
n_f\simeq 0.02M_7^{2/3}(1+z)^3
~{\rm cm^{-3}},
\end{equation}
and
\begin{equation}
t_J=5\times 10^8M_7^{-1/3}(1+z)^{-3/2}~{\rm yr},
\end{equation}
so that for star formation efficiency $\varepsilon>0.03$ charactersitsic time 
$t_{\rm sf}$ remains shorter that the Hubble time for the halo masses 
$M_h>10^7~M_\odot$. This means that star formation rate in merging halos is 
determined by the longest time -- charactersitic time between subsequent 
mergings. In these conditions star formation rate is proportional to the merger 
rate of the halos with the total mass above a critical value $M_{\rm cr}(z)$ 
\cite{barkana00,santos} 

\begin{equation}
\label{sfrate}
\dot M_\ast={1\over2}{\Omega_b\over\Omega_m}\varepsilon f_\perp 
\int\limits_{M_0}^{M_{crit}}dM_1 F(M_1)
\int\limits_{M_{crit}}^{M_{crit}+M_1}dM_2 f_c(M_2,z)M_2 {d^2P\over dM_2 dt},
\end{equation}
where $M_0$ is the minimum mass of a smaller (absorbed) halo,
$P = P(M_1,M_2,z)$ is the probability that
a halo with mass $M_1$ merges with a halo of mass $M_2>M_1$ at redshift $z$
\cite{lacey}; we explicitly assume here that only fraction of baryons 
$f_c(M,z)$ cooled down after merging is able to form stars. Therefore, if 
we substitute here the fraction $f_c$ of baryons cooled to $T<150$ K 
as shown in Fig. 2, 
equation (\ref{sfrate}) will describe the contribution to the total 
star formation rate from the halos where thermodynamics of star forming mergers 
is controlled by HD cooling, and where low mass stars can, in principle, form. 
The critical mass for such mergings can be identified with the mass 
$M_{\rm HD}$, where $f_{c,{\rm HD}}\geq 0.01$. 
In Fig. 5 this contribution is shown by the dotted line. 
Mergings of halos with mass $M_0\ll M_{crit}$ with the halo 
of critical (or overcritical) mass, involve obviously too small baryon 
mass fraction into sufficiently strong 
compression where HD molecules can cool gas down to low 
temperature. Moreover, the compressed region deviates significantly from 
planar geometry, and Gilden criterion is not applicable to these conditions 
anymore. Therefore, in our estimates of star formation rate we assumed 
for $M_0$ two values: $M_0=0.5~M_{crit}$ and $M_0=0.9~M_{crit}$. 
In addition we introduced in (\ref{sfrate}) factor $f_\perp=0.05$ accounting 
only approximately head-on collisions. Indeed, 
our conclusions about the role of HD cooling 
are based on the assumption of a head-on
collision of merging halos, and can be valid only in a restricted range of the
impact parameter when the shear motion is less important than the converging flow
and the corresponding diverging shock waves. For this condition to be 
fulfilled the
characteristic time of the Kelwin-Helmholtz instability of the shear flow
$t_{_{\rm KH}}\sim R/v_{||}$
must be longer than the dynamical (crossing) time $t_{d}\sim 2R/v_\perp$,
where $v_{||}$ and $v_\perp$ are the relative velocity component parallel and
perpendicular the contact discontinuity: $t_{_{\rm KH}}>t_{d}$. This gives
$v_{||}/v_\perp<1/2$, and as a result, only a fraction $f_\perp=
\Delta \Omega/4\pi\simeq 0.05$ of mergers where the flows are approximately head-on.
With this proviso, the two cases: $M_0=0.5~M_\odot$ and $M_0=0.9~M_\odot$, 
are shown in Fig. 5 by dotted and dot-dashed lines, respectively -- 
the region between the two lines can be reasonable estimate of the star formation 
rate where thermodynamics of baryons is dominated by HD cooling.
For comparison we add two lines corresponding to the critical mass of a halo with
$T_{vir} = 400, 10^4$~K (see, Barkana \& Loeb 2000). 
It is obvious, that the number of mergers where HD cooling dominates, is only 
a small but not negligible fraction of all mergers: 
at $z$ between 10 and 16 HD dominated star formation varies from 10 to 
30 \% of the one connected with 10$^4$~K halo mergers (dashed line), and 
from 3 to 20 \% of star formation in 400~K mergers; 
in Fig. 5 for all low-mass mergers dominated by H$_2$ cooling 
$f_c(M,z)=f_{c,{\rm H}_2} = 0.08$ has been assumed following \cite{abel98}. 
Note, however, that at earlier stages, $z=18-20$, mergers with a predominance 
of HD cooling contribute less than 0.5\% compared to the 400~K mergers. 
From this point of view one can expect that in numerical simulations the 
regions with HD cooling can be missed.

It is therefore seen that a small fraction of baryons in mergers cools 
down to the lowest possible temperature $T\simeq T_{\rm CMB}$ and can give 
rise to formation of the first generation stars of low masses -- lower than 
the masses formed under the 
conditions when H$_2$ cooling controls thermal evolution of baryons. 
This fraction increases with the total mass of merging halos, and 
therefore in massive galaxies the population of low-mass first generation 
stars can be considerable. One should stress though that massive galaxies, 
formed in the hierarchical scenario through mergers of less massive systems, 
are quite expected to have been experienced already star formation episodes with 
the insterstellar gas polluted by metals. However, it remains unclear whether 
the metals can become well mixed in a galaxy before it absorbs a new halo 
in next merger event, and simple arguments suggest the opposite. 
Indeed, the characteristic mixing time for the whole galaxy 
can be estimated as $t_{mx}\sim \langle \delta_{ej}\rangle R/c_s$, where 
$\langle \delta_{ej}\rangle>1$ is mean density contrast between SNe ejecta 
and diffuse interstellar gas, $R$ is the galaxy radius, $c_s$ is the sound speed 
or the velocity dispersion in diffuse gas; note that $R/c_s\sim t_c$.
The characteristic time scale for the halo mass 
growth $t_m = [M_2 d^2 P(M_1,M_2,z)/dM_2dt]^{-1}$, 
where $P(M_1,M_2,z)$ is the 
probability that a halo of mass $M_2$ absorbs a smaller halo $M_1$ 
\cite{lacey}. For the halos
$M_2 \sim 10^7~M_\odot$ and $M_1\sim 0.9 M_2$ at $z =20$ $t_m$ is about 
$\sim 8\times 10^{14}~$s, which is comparable to the collision time
$t_c = 3R/2v_c$, and is therefore $<t_{mx}$. Although some of the absorbed 
low mass halos can have been experienced star formation episodes before 
being merged, and thus can be alredy metal enriched \cite{ferraraxx}, 
however a non-negligible fraction of them may have pristine composition. 
\cite{cen06} studied the star formation history before reionization
and found that the era of Pop III star formation can be significantly 
prolonged. From this point of view one can expect possible important 
contribution from primordial star formation at redshifts 
$z \sim 3-4$ \cite{haiman}.

The existence of low mass Pop III stars is suspected from the 
observational point of view: recently discovered extremely metal-poor 
low-mass stars, as for instance a $0.8~M_\odot$ star 
HE 0107-5240 \cite{christ02,christ04}
with [Fe/H]$=-5.3$, may be the first-generation stars. The question of whether 
stars with [Fe/H]$<-5$ are indeed Population III stars is still 
under discussion, partly because of overabundant carbon and nitrogen: 
[C/Fe]$=4$, [N/Fe]$=2.3$ in HE 0107-5240 \cite{bessel04}. 
From this point of view HE 0107-5240 can be a Population II star formed in 
already enriched interstellar gas \cite{umeda03}. 
However, the possibility that this star has formed of pristine gas cannot be 
excluded: \cite{shige03} conclude that HE 0107-5240 is a Pop III 
star with the surface polluted by accreted interstellar gas already enriched 
with metals. In this scenario overabundant C and N in the envelope 
can be produced during the core helium flash as suggested 
by \cite{weiss00,schlattl02}. 

\section{Conclusion}

In this paper we have shown that 

\begin{itemize}

\item

A small fraction of baryons in merging halos can cool down to very 
low temperatures close to the temperature of the cosmic microwave 
background; 

\item

This fraction increases with the halo mass, and can reach $\simeq 0.2$ 
for masses $M>2\times 10^7~M_\odot$ at $z=20$ and for 
$M>10^8~M_\odot$ for $z=10$; 

\item

Such cold gas is unstable against gravitational fragmentation, 
with the mass of the primary fragments decreasing for mergers of higher 
masses $M$: $M_J\leq 1.3\times 10^4M^{-1/3}M_\odot$; masses of protostars 
formed in gas cooled down by HD molecules are in general lower than those 
formed in conditions when H$_2$ cooling dominates;

\item

The contribution to the cosmic star formation rate of the mergers with a 
predominance of HD cooling, and therefore with presumably low mass first 
stellar objects, increases from less than 0.5\% at redshift $z=18-20$ to 
10-30 \% at $z=10$. Extremally metal-poor low mass stars in the 
Milky Way may have been formed in mergers dominated by HD cooling. 

\end{itemize}

\section{Acknowledgments}

EV thanks E. Ripamonti for useful discussions. This work is 
supported by the Federal Agency of Education, project code RNP 2.1.1.3483, 
by the RFBR (project code 06-02-16819-a) 
and by the Rosnauka Agency grant No 02.438.11.7001.

\newpage

\begin{figure*}
  \vspace{24pt}

\includegraphics[width=110mm]{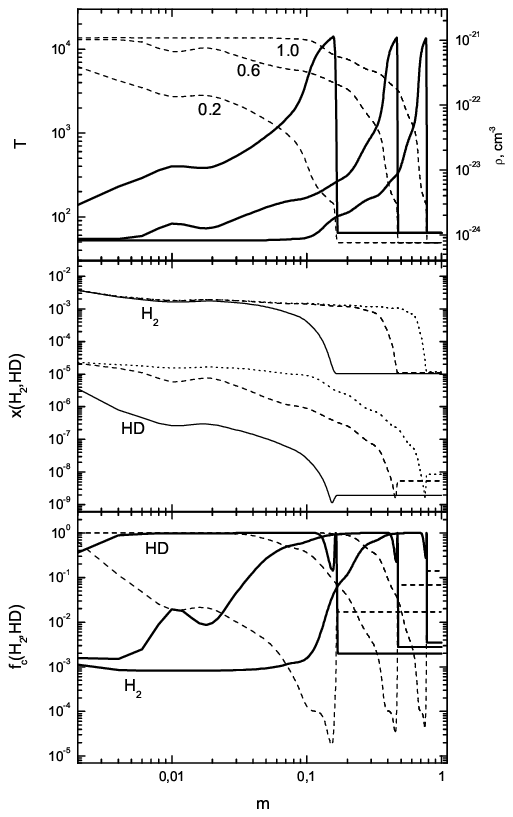}
      \caption{Upper panel: profiles of temperature (solid) and density (dash);
      middle panel: relative concentration of H$_2$ and HD molecules; 
      lower panel: their relative contribution to the total cooling
      (H$_2$ -- solid and HD -- dashed), for 
      baryons in two colliding halos with the total mass $M=1.9\times 10^7M_\odot$
      at $0.2t_c,~0.6t_c,~t_c$; halos merged at $z=20$.
                }
         \label{Fig1}
\end{figure*}

\begin{figure*}
  \vspace{24pt}

\includegraphics[width=130mm]{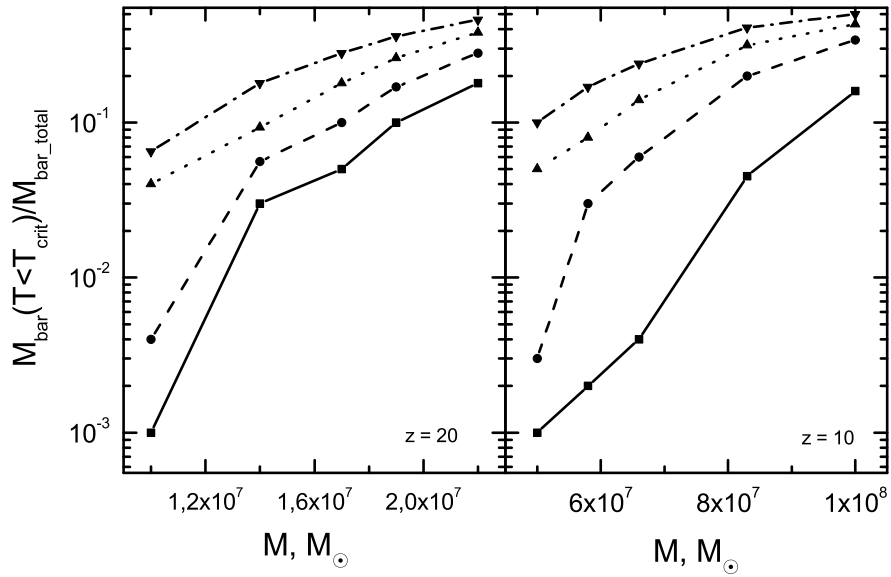}
      \caption{Fraction of baryons cooled below temperature $T<200$ K,
      $T<150$ K, $T<100$ K, and $T=T_{\rm CMB}$ from top to bottom at 
      $z = 20$ (left), and $z = 10$ (right).
                }
         \label{Fig2}
\end{figure*}

\begin{figure*}
  \vspace{24pt}

\includegraphics[width=120mm]{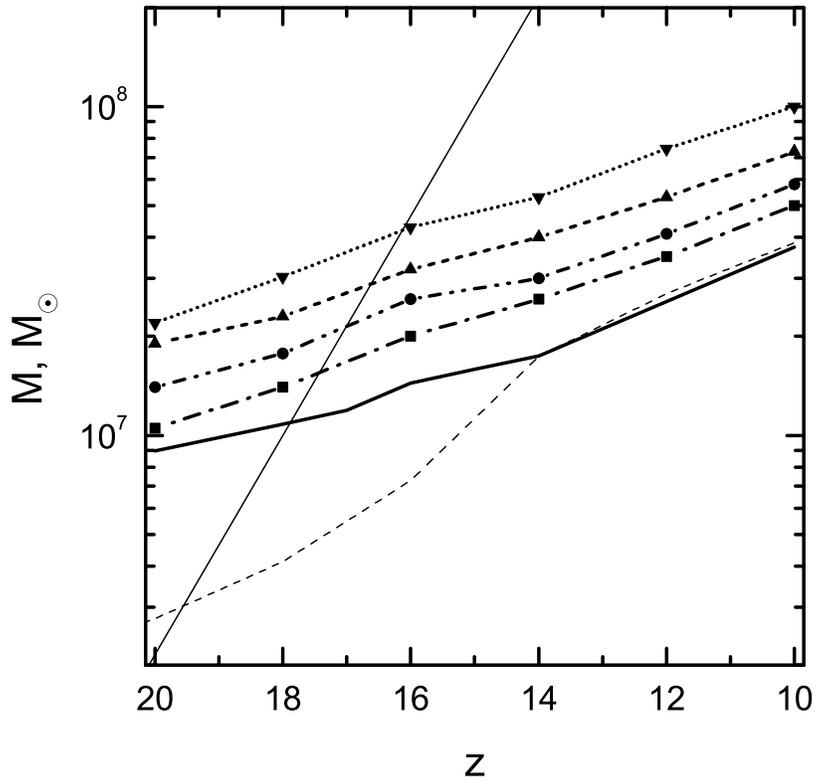}
      \caption{Lines with simbols depict halo masses vs redshift, where
      the fraction of baryon mass unstable in Gilden sense is at given level:
      $1\%,\ 7\%,\ 25\%,\ 50\%$ from bottom to top; straight thin solid line
      corresponds to a 3$\sigma$ peak mass, dashed line shows the minimum
      mass obtained by (Tegmark et al., 1997), thick solid lines shows the 
      minimum mass from (Shchekinov \& Vasiliev 2006). 
      					}
         \label{Fig3}
\end{figure*}

\begin{figure*}
  \vspace{24pt}

\includegraphics[width=120mm]{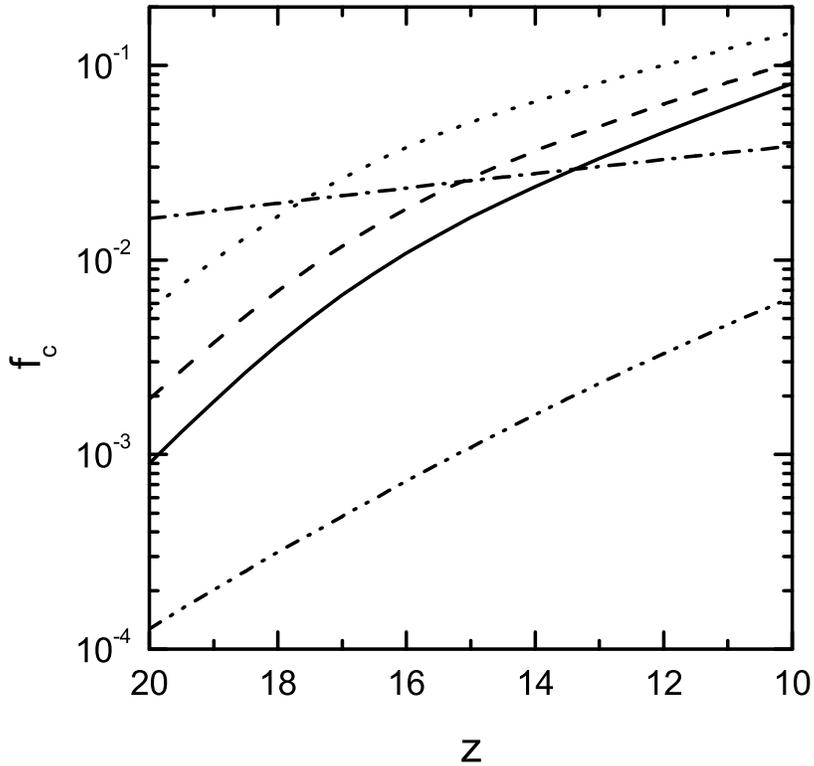}
      \caption{Fraction of baryons $f_c$ in 
        the universe cooled below 100 K. According
        to Gilden criterion these baryons can give rise to the formation of
        stars. Solid curve shows the fraction $f_c$ for $M_{\rm min}=
        10^3M_\odot$ in eq. (\ref{massint}), dashed curve -- 
        $M_{\rm min}=10^4M_\odot$, dotted  --  $M_{\rm min}=10^5M_\odot$
        dot-dashed and dot-dot-dashed curves correspond to $M_{\rm c}$ 
        with $T_{vir} = 400, 10^4$~K}.
         \label{Fig4}
\end{figure*}

\begin{figure*}
  \vspace{24pt}

\includegraphics[width=120mm]{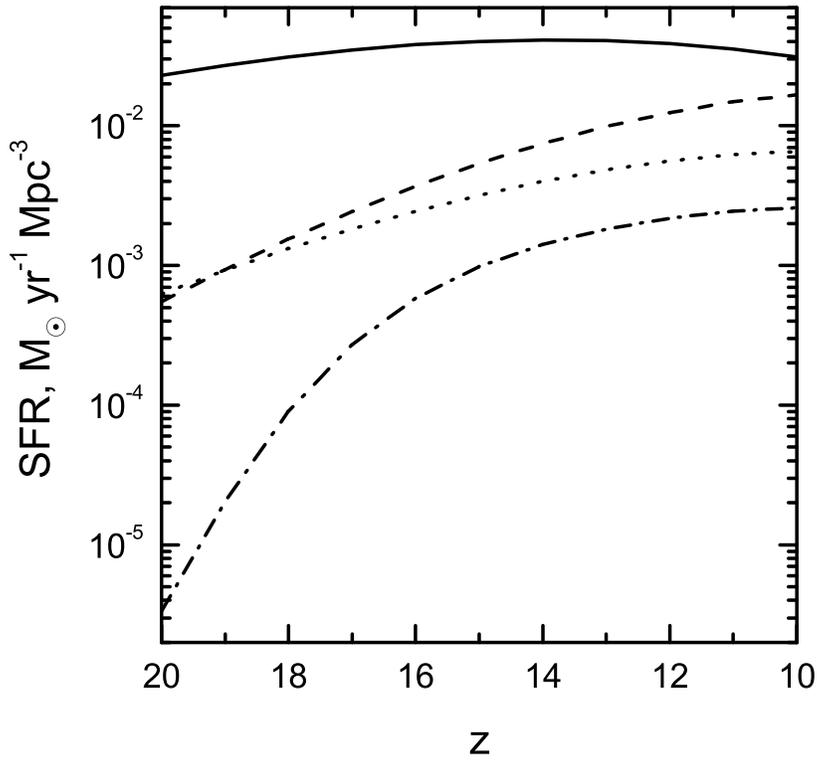}
      \caption{Cosmic star formation rate: 
      solid and dashed lines correspond to the halos with
      $T_{vir} = 400, 10^4$~K; 
      dotted -- the halos with mass $M_{crit} = M_{\rm HD}$, where 
      $f_{c,{\rm HD}}\sim 0.01$ and $M_0 = 0.5M_{crit}$, and 
      dot-dashed line -- $M_0 = 0.9M_{crit}$
      (see text for details); in all cases $\varepsilon=0.1$.}
         \label{Fig5}
\end{figure*}

\end{document}